\documentclass[aps,prd,twocolumn,floatfix,superscriptaddress,longbibliography,10pt]{revtex4-2}

\usepackage[latin9]{inputenc}
\usepackage{amsmath}
\usepackage{amssymb}
\usepackage{graphicx}
\usepackage{esint}
\usepackage{epstopdf}
\usepackage{braket}
\usepackage{bm}
\usepackage[dvipsnames]{xcolor}
\usepackage[colorlinks=true,citecolor=blue,linkcolor=blue,urlcolor=blue]{hyperref}
\usepackage{bibunits}
\usepackage{mathtools}
\usepackage{stackengine}
\usepackage{dsfont}
\usepackage{soul}
\usepackage{multirow}
\usepackage{placeins}
\usepackage{xspace}
\usepackage{nicefrac}

\newcommand{\ttbar}{\ensuremath{\text{t} \bar{\text{t}}}\xspace}

\newcommand{\ttm}{\ensuremath{m(\ttbar)}\xspace}
\newcommand{\abs}[1]{\ensuremath{\lvert #1 \rvert}}

\newcommand{\gluglu}{\ensuremath{\text{gg}}\xspace}
\newcommand{\qqbar}{\ensuremath{\text{q}\bar{\text{q}}}\xspace}
\newcommand{\qglu}{\ensuremath{\text{q}\text{g}}\xspace}

\newcommand{\thetat}{\ensuremath{\theta}\xspace}
\newcommand{\costh}{\ensuremath{\cos(\thetat)}\xspace}
\newcommand{\absct}{\ensuremath{\abs{\costh}}\xspace}

\newcommand {\ie}{\mbox{i.e.}\xspace}

\newcommand {\vs}{\mbox{\textsl{vs.}}\xspace} 

\newcommand{\ttmcth}{\mbox{\ttm \vs \absct}\xspace}

\newcommand{\POWHEG} {{\textsc{powheg}}\xspace}

\newcommand{\PYTHIA} {{\textsc{pythia}}\xspace}

\newcommand{\PWGPYT}{\POWHEG{}+\PYTHIA}

\newcommand{\Mtwo}{\ensuremath{\tilde{\mathcal{M}}_\mathrm{2}}\xspace}
\newcommand{\steer}{\ensuremath{\mathcal{T}}\xspace}

\newcommand{\rtb}{\ensuremath{ \rho_{\bar{\text{t}}} }\xspace}
\newcommand{\rttbar}{\ensuremath{\rho_{\ttbar}}\xspace}

\newcommand{\biv}{\ensuremath{\mathcal{B}}\xspace}

\newcommand{\DE}{\ensuremath{\Delta_\mathrm{E}}}
\newcommand{\qd}{\ensuremath{\mathcal{D}_{\text{t}}}\xspace}
\newcommand{\qdb}{\ensuremath{\mathcal{D}_{\bar{\text{t}}}}\xspace}

\begin{document}

\title{Experimental characterization of the hierarchy of quantum correlations in top quark pairs}

\author{Yoav Afik}
\email{yoavafik@gmail.com}
\affiliation{Enrico Fermi Institute, University of Chicago, Chicago, Illinois 60637, USA}

\author{Regina Demina}
\email{regina@pas.rochester.edu}
\affiliation{Department of Physics and Astronomy, University of Rochester 500 Joseph C. Wilson Boulevard, Rochester, NY 14627, USA}

\author{Alan Herrera}
\email{alan.herrera@cern.ch}
\affiliation{Department of Physics and Astronomy, University of Rochester 500 Joseph C. Wilson Boulevard, Rochester, NY 14627, USA}

\author{Otto Heinz Hindrichs}
\email{otto.heinz.hindrichs@cern.ch}
\affiliation{Department of Physics and Astronomy, University of Rochester 500 Joseph C. Wilson Boulevard, Rochester, NY 14627, USA}

\author{Juan Ram\'on Mu\~noz de Nova}
\email{jr.denova@csic.es}
\affiliation{Instituto de Estructura de la Materia, IEM-CSIC, Serrano, 123 E-28006 Madrid, Spain}
\affiliation{Departamento de F\'isica de Materiales, Universidad Complutense de Madrid, E-28040 Madrid, Spain}

\author{Baptiste Ravina}
\email{baptiste.ravina@cern.ch}
\affiliation{CERN, CH-1211 Geneva, Switzerland}

\begin{abstract}
Recent results from the Large Hadron Collider have demonstrated quantum entanglement of top quark--antiquark pairs using the spin degrees of freedom. 
Based on the doubly differential measurement of the spin density matrix of the top quark and antiquark performed by the CMS collaboration in the helicity and beam bases, we evaluate a set of quantum observables, 
including discord, steerability, Bell correlation, and magic. 
These observables allow for a quantitative characterization of the quantum correlations present in a top quark--antiquark system, thus enabling an interpretation of collider data in terms of quantum states and their properties.
Discord is observed to be greater than zero with a significance of more than 5 standard deviations ($\sigma$) in several regions of phase space, some of which correspond to separable quantum states. Evidence for steerability is established for the first time in a high-energy system, with a significance of more than 3$\sigma$. No Bell correlation is observed within the currently probed phase space, in agreement with the theoretical prediction. These results experimentally corroborate the hierarchy of quantum correlations in top quarks with discord being the most basic form of quantum correlation, followed by entanglement, steerability, and Bell correlation. 
The significance of nonzero magic, which is a complementary observable to the quantum correlation hierarchy, is found to exceed 5$\sigma$ in several regions of phase space. 
\end{abstract}

\maketitle

\tableofcontents

\section{Introduction}

Quantum mechanics (QM) underpins both quantum information science (QIS) and high-energy physics (HEP). 
Despite this shared foundation, the two fields differ significantly in their objectives and have, over time, developed distinct conceptual frameworks and terminologies. 
While HEP is primarily concerned with understanding the fundamental constituents and forces of nature, QIS is focused on harnessing quantum principles to develop powerful computational and communication technologies. Deepening our understanding of the foundational aspects of QM can lead to further progress in QIS. Although the Large Hadron Collider (LHC) was not originally designed for these studies, it can still serve as a unique laboratory for probing these aspects in a high-energy regime inaccessible elsewhere. To effectively use HEP systems for this purpose, it is important to establish a bridge between QIS and HEP. 
In this work, based on experimental results, we establish a correspondence between both fields by evaluating a number of fundamental observables common in QIS within a HEP system, in particular top quarks.

Top quark--antiquark pairs (\ttbar) are produced at the LHC predominantly via gluon-gluon fusion (\gluglu), with additional contributions from quark-antiquark annihilation (\qqbar). Processes involving (anti)quark-gluon initial states (\qglu) contribute through higher-order QCD corrections.
Top quarks decay almost always to a $W$ boson and a $b$ quark~\cite{ParticleDataGroup:2024cfk}. Typically, the \ttbar system is characterized in terms of the decays of the two $W$ bosons. When both of them decay to a charged lepton and a neutrino, the event is referred to as dilepton. If one $W$ boson decays leptonically, while the other one decays hadronically to a $d$-type ($d$ or $s$) and a $u$-type ($u$ or $c$) quark, the channel is called lepton+jets, since quarks hadronize into jets.

As the most massive fundamental particle known, with a mass of $m_t \approx 172.5$~GeV, the top quark has an exceptionally short lifetime ($\tau_t \approx 5 \times 10^{-25}$~s), which is far shorter than the characteristic timescales for hadronization ($1/\Lambda_{\rm QCD} \approx 3 \times 10^{-24}$~s) and spin decorrelation ($m_t/\Lambda_{\rm QCD}^2 \approx 2 \times 10^{-21}$~s)~\cite{Bigi:1986jk,ParticleDataGroup:2024cfk}.
Consequently, the spin information of the top quarks is transmitted directly to their decay products, and their angular distributions can be used to extract the polarizations and spin correlations of a \ttbar pair~\cite{Bernreuther1994,Bernreuther1998,Bernreuther2001,Bernreuther:2004jv,Bernreuther2015}. 

Based on these properties, various methods were put forward to measure quantum correlations in the \ttbar system, such as discord~\cite{Afik:2022dgh,Han:2024ugl}, entanglement~\cite{Afik:2020onf}, steerability~\cite{Afik:2022dgh}, and Bell correlation~\cite{Fabbrichesi:2021npl,Afik:2022kwm,Severi:2021cnj,Aguilar-Saavedra:2022uye,Dong:2023xiw,Han:2023fci,Cheng:2024btk}. Interestingly, these correlations follow a logical hierarchy~\cite{Steer_epr,Qureshi2018,Baker2020,Uola2020}, where the presence of Bell correlations implies steerability, which in turn implies entanglement, and this implies discord. Beyond quantum correlations, there are a number of alternative genuine quantum features (such as coherence~\cite{Haddadi:2026nyg}, contextuality~\cite{Fabbrichesi:2025rsg}, quantum predictability~\cite{Moreno2026}, or magic~\cite{White:2024nuc}) that can be studied in top quarks, reflecting various intriguing manifestations of QM and complementing the picture provided by the quantum correlation hierarchy. For a non-exhaustive review of the possibility to measure QIS observables in colliders, see Refs.~\cite{Barr:2024djo,Afik:2025ejh}. 

Entanglement in the \ttbar system was observed by the ATLAS~\cite{ATLAS:2023fsd} and CMS~\cite{CMS:2024pts,CMS:2024zkc,CMS:2025brx} collaborations at the LHC. These results demonstrated entanglement at the highest energy ever probed. Based on these data, it was reported that in some regions of phase space, the spin states of the top quark and antiquark are separable, yet they exhibit nonzero discord and magic~\cite{Fabbrichesi:2025psr}. Nevertheless, the evaluation of the quantum-correlation hierarchy from the experimental data is still lacking in the literature. Based on the CMS doubly differential measurement of the \ttbar spin density matrix, performed in the lepton+jets channel in the helicity and beam bases at $\sqrt{s}=13$~TeV~\cite{CMS:2025brx,CMS:2024zkc}, we evaluate quantum discord, steerability, Bell correlation, and magic.

\section{Theoretical framework}
\subsection{\ttbar system at the LHC}

The top quark has a spin of $\nicefrac{1}{2}$ and, therefore can be represented as a qubit, the most basic unit of information in QIS. Hence, \ttbar can be represented by a two-qubit system, with its density matrix expanded as 
\begin{equation}
\begin{aligned}
    \rttbar =  \frac{1}{4} ( I_2 \otimes I_2 & + \sum_i P_i \sigma_i \otimes I_2 \\ & + \sum_j \bar{P}_j I_2 \otimes \sigma_j + \sum_{ij}  C_{ij} \sigma_i \otimes \sigma_j   )~,
    \end{aligned}
\label{eq_densitymatrix}
\end{equation}
where $I_2$ is the $2\times 2$ identity matrix, $\sigma_i$ are Pauli matrices, $P_i$ and $\bar{P}_j$ describe the net polarization of the top quark and antiquark, respectively, and $C_{ij}$ represents their spin correlation matrix. 
The 3-component vectors $P_i$, $\bar{P}_j$ and the $3\times 3$ matrix $C_{ij}$ are extracted from the angular distribution of the \ttbar decay products with the highest spin analyzing power ($\approx\!1$)---charged leptons and $d$-type quarks~\cite{Brandenburg:2002xr}. 
The indices $i$ and $j$ correspond to the axes of the basis selected for the measurement.

Common choices for the spin measurement are the beam and helicity bases. 
Both are right-handed coordinate systems defined in the \ttbar center-of-mass frame~\cite{Bernreuther:2004jv}. The beam basis is defined by the unit vectors $({\hat{x}, \hat{y}, \hat{z}})$, with the $x$ axis pointing toward the center of the LHC ring, the $y$ axis oriented vertically upward, and the $z$ axis pointing along the proton beam axis, $\hat{z}=\hat{p}$~\cite{ATLAS:2008xda,CMS:2008xjf}. The helicity basis~\cite{Baumgart:2012ay} is defined by the unit vectors $({\hat{r}, \hat{k}, \hat{n}})$, where $\hat k$ denotes the flight direction of the top quark, $\hat{r}=(\hat{p}-\cos\theta\,\hat{k})/\sin\theta$ and $\hat{n}=\hat{r}\times\hat{k}$. Here, $\theta$ is the top quark production angle with respect to $\hat{p}$, with $\cos\theta=\hat{k}\cdot\hat{p}$. For the helicity basis we use the sign convention of Ref.~\cite{Bernreuther2015},
\begin{equation}
\hat n \to \mathrm{sgn}(\cos\theta) \cdot \hat n,~~~~~\hat r \to  \mathrm{sgn}(\cos\theta) \cdot \hat r,~~~~~\hat k \to \hat k~.
\end{equation}
The evaluation of the different quantum observables considered here depends on the chosen basis. Specifically, in an event-dependent basis, such as the helicity basis, the axis orientation is modified according to the underlying \ttbar kinematics, leading, after integration over some phase-space variables (e.g., an azimuthal angle), to what has been referred to as fictitious quantum states~\cite{Afik:2020onf, Afik:2022kwm, Cheng2023, Cheng:2024btk}. On the other hand, the beam basis keeps its orientation fixed, leading to \textit{bona fide} quantum states.

In a two-qubit system there are four maximally entangled states, referred to as Bell states: $\ket{\Phi^{\pm}}=\frac{1}{\sqrt 2} (\ket{\uparrow \uparrow} \pm \ket{\downarrow \downarrow})$ and $\ket{\Psi^{\pm}}=\frac{1}{\sqrt 2} (\ket{\uparrow \downarrow} \pm  \ket{\downarrow \uparrow})$, where the arrows indicate here the spin orientation with respect to the $\hat k$ axis. 
In Ref.~\cite{Afik:2020onf} it was demonstrated that at the threshold of \ttbar production (i.e., when the invariant mass of the \ttbar system, \ttm, is close to twice the mass of the top quark), the system is predominantly in a $\ket{\Psi^-}$ spin-singlet state. 
The observed entanglement in the dilepton channel~\cite{ATLAS:2023fsd, CMS:2024pts} corresponds to this region of the \ttbar phase space. Notably, in the threshold region, the decays of the top quark and antiquark are separated predominantly by a time-like interval~\cite{Severi:2022qjy, Demina:2024dst, Fabbrichesi:2025psr}.
The lepton+jets channel is better suited for measurements in the higher \ttm region~\cite{CMS:2024zkc, CMS:2025brx}, where the system is expected to be predominantly observed in a $\ket{\Phi^-}$ spin-triplet state~\cite{Afik:2020onf}. 
Indeed, in Ref.~\cite{CMS:2025brx} it was shown that the \ttbar system in the $\ttm >800$~GeV, $\absct <0.4$ bin is observed in the $\ket{\Phi^-}$ state with about 60\% fidelity, the highest value observed with respect to the Bell basis in the probed phase space.

The entanglement marker  
\begin{equation}
\DE\equiv-C_{33}+|C_{11}+C_{22}|,
\label{eq:EntanglementMarker}
\end{equation}
with $\DE>1$ being a sufficient condition for entanglement \cite{Afik:2020onf}, can be derived from the Peres--Horodecki criterion~\cite{Peres:1996dw, Horodecki:1997vt}, where the indices $(1,2,3)$ correspond to $(r,k,n)$ and $(x,y,z)$ in the helicity and beam basis, respectively. Using this entanglement marker, it was demonstrated that the \ttbar system is in an entangled state in the high-\ttm region of phase space~\cite{CMS:2024zkc}. In contrast to the threshold, in this region the decays of the top quark and antiquark are separated by a space-like interval~\cite{Severi:2022qjy, Demina:2024dst, Fabbrichesi:2025psr}. 

Here, we use the results of the doubly differential measurement in bins of \ttmcth of the full spin correlation matrix reported in Ref.~\cite{CMS:2024zkc} for the helicity basis and in Ref.~\cite{CMS:2025brx} for the beam basis. 
In both cases, we have found good agreement between the measured spin correlation coefficients and the ones obtained analytically~\cite{Afik:2020onf,Afik:2022kwm}. 

\subsection{Discord}

In classical information theory, the mutual information of two systems characterizes how much information is obtained about one of them by observing the other. One of the hallmark signatures of QM is the difference between two classically equivalent expressions for the mutual information, quantified by the so-called quantum discord~\cite{Ollivier2001}. For a two-qubit \ttbar system, the discord for the top quark reads~\cite{Hamieh2004,Lu2011,Girolami2011} 
\begin{align}\label{eq:discordA}
\qd 
&=  S(\rtb) - S(\rttbar) \nonumber\\
&+ \min_{\hat u}\Big[
p_{t,+\hat u} S(\rho_{t,+\hat u})
+ p_{t,-\hat u} S(\rho_{t,-\hat u})
\Big]~,
\end{align}
where $S(\rho) = -\mathrm{Tr}(\rho \log_2 \rho)$ is the von Neumann entropy, $\rho_{t}=\mathrm{Tr}_{\bar{t}}[\rttbar]$ is the reduced density matrix for the top quark, $\rho_{\bar{t}}=\mathrm{Tr}_{t}[\rttbar]$ is the reduced density matrix for the top antiquark, and $\rho_{t,\pm \hat u}$ represent the post-measurement quantum states for the top quark given a projective spin measurement on the top antiquark along a certain quantization axis $\hat u$, with outcomes characterized by the projectors $\Pi_{\pm \hat u}$. Specifically, 
\begin{equation}
\rho_{t,\pm \hat u}=\frac{1}{p_{\pm \hat u}}\,
\mathrm{Tr}_{\bar t}\!\left[ (I_2\otimes\Pi_{\pm \hat u})~\rttbar (I_2\otimes\Pi_{\pm \hat u}) \right]~,
\end{equation}
where $p_{t,\pm \hat u}$ are the probabilities of each outcome,
\begin{equation}
p_{t,\pm \hat u} = \mathrm{Tr}\!\left[(I_2\otimes\Pi_{\pm \hat u})~\rttbar (I_2\otimes\Pi_{\pm \hat u}) \right]~.
\end{equation}
Finally, the minimization over the directions of $\hat u$ in Eq.~\eqref{eq:discordA} represents the minimization over all possible projective measurements, ensuring the least disturbance on the system by the observation process~\cite{Ollivier2001}. In this work, this minimization is performed numerically. A similar expression can be derived for the discord of the top antiquark.

\subsection{EPR paradox, steering, and Bell's theorem }

In their famous 1935 paper, Einstein, Podolsky and Rosen (EPR) questioned the completeness of QM by highlighting a paradox---an apparent incompatibility between quantum predictions and the principles of locality and realism~\cite{EPR_paradox}. 
EPR considered a pair of spatially separated systems prepared in a correlated quantum state, such that the measurement of one system enables a prediction, with certainty, of the outcome of a corresponding measurement on the other. 
Under the assumption that physical influences cannot propagate superluminally, EPR argued that these correlations imply the existence of elements of reality (hidden variables) not captured by the quantum description, concluding that QM must be incomplete. 

The concept of quantum steering, defined as the ability to influence, or ``steer", the quantum state of one system by performing a measurement on another, arises naturally from the EPR argument. Schr{\"{o}}dinger proposed the concept of steering as his interpretation of the EPR paradox~\cite{Schrodinger_1935}, but the precise formulation was given much later in Ref.~\cite{Steer_epr}. A two-qubit system in an unpolarized state (i.e., $P_i=\bar{P}_j=0$), such as \ttbar at leading order in QCD~\cite{Afik:2022dgh}, is steerable if and only if~\cite{Jevtic:2015epl,Nguyen2016}
\begin{equation}
    \steer \equiv \int \mathrm{d}^2\hat u\,\sqrt{\hat u^{T} C^{T} C\,\hat u} > 2\pi~,
    \label{eq_steering}
\end{equation}
where the integral is taken over all possible orientations of the unit vector $\hat u$. 

The formulation of the EPR paradox initiated a sustained experimental and theoretical effort to characterize nonclassical correlations. A Bell inequality sets an upper bound, which must be respected by local hidden-variable theories (LHVTs), on the correlations between measurement outcomes on two subsystems. Bell's theorem~\cite{Bell:1964kc} shows that, while all LHVTs must satisfy these inequalities, the correlations predicted by QM can violate them. Experimental tests have indeed observed such violations~\cite{Aspect1982,Hagley1997,Belle:2007ocp,Storz:2023jjx}, thereby ruling out LHVTs and supporting the QM description of nature.

In a high-energy collider environment, Bell correlations are evaluated via quantum tomography of the underlying two-qubit state. This strategy does not constitute a direct test of Bell inequality and, thus, cannot exclude LHVTs~\cite{Abel:2025skj,Bechtle:2025ugc, Abel:1992kz}. Instead, it assesses whether the reconstructed state would be expected to violate a standard Bell test.

A useful form of Bell inequality is the Clauser--Horne--Shimony--Holt (CHSH) inequality~\cite{Clauser1969}, which is violated in a two-qubit system if and only if the two largest eigenvalues of $C^TC$ ($m_1$ and $m_2$) satisfy~\cite{Horodecki1995, Fabbrichesi:2021npl,Afik:2022kwm}
\begin{equation}
    \biv \equiv m_1+m_2>1~.
    \label{bell_ineq}
\end{equation}
A state fulfilling this criterion is referred to as a Bell-correlated state~\cite{Brunner:2013est}. The experimental observation of such a state at the LHC is of great interest~\cite{Low:2025aqq,Fabbrichesi:2025aqp}, since it would demonstrate that the most extreme manifestations of quantum behavior can also be achieved in high-energy colliders. 

It is known that quantum correlations follow a well-structured hierarchy~\cite{Steer_epr,Qureshi2018,Baker2020,Uola2020}, with discord being the most basic one, followed by entanglement, steerability, and then Bell correlation.

\subsection{Magic}

Quantum computers promise significant advantages over classical computation~\cite{Nielsen2000}. Key quantum phenomena such as superposition and entanglement enable quantum algorithms to achieve exponential or polynomial speedups for specific classes of problems compared to their classical counterparts~\cite{Shor:1994jg,Shor:1994ihq,Grover:quantalgo, Grover:1997fa}. However, the Gottesman--Knill theorem~\cite{GottesmanKnill:1998hu} states that quantum circuits involving only stabilizer states, defined as the simultaneous eigenstates of a set of commuting Pauli operators, and operations from the Clifford group can be efficiently simulated on a classical computer. For example, Shor's algorithm~\cite{Shor:1994jg}, one of the most prominent examples of quantum advantage and a milestone in modern cryptography, is based on non-Clifford gates, and offers polynomial-time solutions to integer factorization, a task for which the best known classical algorithms have sub-exponential complexity.
Quantum magic is a property of quantum states that quantifies their computational advantage over classical resources, with stabilizer states presenting zero magic.
Magic is also studied in broader contexts such as black hole dynamics, where the growth of their interiors cannot be explained by entanglement on the boundary alone~\cite{Goto:2021anl}, and in other fields such as quantum chaos, many-body theory, and in potential simulations of quantum gravity~\cite{Leone2021quantumchaosis, White:2020zoz, Liu:2020yso, Gu:2024ure, Oliviero:2022euv,Rattacaso:2023kzm, Tarabunga:2023ggd, Zhou:2020scv, Gu:2024qvn, Cepollaro:2024qln}. 

A general form of magic \Mtwo for mixed quantum states, such as the \ttbar system, is based on the second stabilizer R\'enyi entropy~\cite{RenyiEntropy:2022,White:2024nuc} and can be evaluated as
\begin{equation}
    \Mtwo=-\log_{2} \left(  \frac{1+ \sum_{i}[ (P_i^4 + \bar{P}_i^4) ] + \sum_{i,j}C_{ij}^4}{1+ \sum_{i}[ (P_i^2 + \bar{P}_i^2) ] + \sum_{i,j}C_{ij}^2}  \right)~.
    \label{eq_1magic}
\end{equation}

\section{Results}
We evaluate quantum discord using Eq.~(\ref{eq:discordA}), steerability using Eq.~(\ref{eq_steering}), Bell correlations using Eq.~(\ref{bell_ineq}), and quantum magic using Eq.~(\ref{eq_1magic}).
The confidence intervals are estimated from the negative logarithm of the likelihood function $L$, evaluated as 
\begin{equation}
    -2\log(L) = \sum_{i,j} (o_i - x_i) U_{ij}^{-1} (o_j - x_j)~,
    \label{eq_lhood}
\end{equation}
where $o_i$ are the observed values of the polarization and spin correlation coefficients in a given \ttmcth bin, and $x_i$ are the corresponding parameters of the likelihood, constrained to be within the physically allowed region $-1 \leq x_i \leq 1$ (in practice, these limits are never reached). The correlations between the observed values $o_i$, encoded in the covariance matrix $U_{ij}$, need to be taken into account for a proper evaluation of the uncertainties on the quantum observables. The minimization of Eq.~(\ref{eq_lhood}) is performed under the condition that the observable of interest, i.e., \qd, \steer, \biv, or \Mtwo, is fixed at a certain value. By construction, the global minimum with $-2\log(L) = 0$ for $x_i = o_i$ defines the central values. For each observable, a range around the central value is scanned, minimizing $-2\log(L)$ under the condition that the observable is fixed at the scanned value. The 68\% confidence interval is determined by the scanned values for which $-2\log(L) = 1$. 

Whenever applicable, we also determine the exclusion limits with respect to the threshold values, i.e., zero for discord and magic, $2\pi$ for steerability, and one for Bell correlation, from the values of the minimized $-2\Delta\log L$ at the thresholds. Significances above $5\sigma$ are quoted only to illustrate the hierarchy among the different quantum correlations and should not be interpreted as precise quantitative values.

The results for all observables are presented in Tables~\ref{tab_hel} and \ref{tab_beam} for the helicity and beam bases, respectively. In both tables, we also show the results for the entanglement marker \DE~from Refs.~\cite{CMS:2024zkc,CMS:2025brx}.
\begin{table*}[htbp]
\renewcommand{\arraystretch}{1.3}
\setlength{\tabcolsep}{5pt}
\centering
\caption{QIS observables in helicity basis with their uncertainties, determined by $-2\log(L) = 1$.
The significance (in units of $\sigma$) of the deviation from the null hypothesis, \ie, \qd$=0$, \DE$=1$, \steer$=2\pi$, \biv$=1$, \Mtwo$=0$, is shown in square brackets when it is greater than $3\sigma$. Bins where the \ttbar system is demonstrated to be in an entangled state are highlighted in bold.}
\begin{tabular}{l l l l l l l}
\hline
\multicolumn{1}{c}{$\ttm$ [GeV]} & \multicolumn{1}{c}{$\absct$} & \multicolumn{1}{c}{\qd} & \multicolumn{1}{c}{\DE} & \multicolumn{1}{c}{\steer} & \multicolumn{1}{c}{\biv} & \multicolumn{1}{c}{\Mtwo} \\
\hline
& $[0.0,0.4]$ & $0.147_{-0.121}^{+0.066}$ & $0.71_{-0.25}^{+0.25}$ & $4.35_{-1.19}^{+0.93}$ & $0.51_{-0.18}^{+0.25}$ & $0.47_{-0.11}^{+0.15}[4.6\sigma]$ \\
$[300,400]$ & $[0.4,0.7]$ & $0.16_{-0.11}^{+0.16}$ & $1.37_{-0.38}^{+0.37}$ & $6.2_{-1.4}^{+1.3}$ & $0.85_{-0.24}^{+0.27}$ & $0.61_{-0.22}^{+0.17}[4.4\sigma]$ \\
 & $[0.7,1.0]$ & $0.18_{-0.11}^{+0.11}[3.4\sigma]$ & $1.43_{-0.26}^{+0.26}$ & $6.38_{-0.92}^{+0.98}$ & $0.80_{-0.18}^{+0.17}$ & $0.580_{-0.133}^{+0.093}[4.8\sigma]$ \\
\hline
 & $[0.0,0.4]$ & $0.218_{-0.052}^{+0.080}[9.2\sigma]$ & $0.419_{-0.064}^{+0.063}$ & $3.88_{-0.27}^{+0.28}$ & $0.240_{-0.031}^{+0.037}$ & $0.340_{-0.035}^{+0.037}[>\!\!10\sigma]$ \\
$[400,600]$ & $[0.4,0.7]$ & $0.111_{-0.034}^{+0.042}[5.5\sigma]$ & $0.340_{-0.077}^{+0.076}$ & $3.12_{-0.33}^{+0.33}$ & $0.147_{-0.032}^{+0.038}$ & $0.236_{-0.042}^{+0.045}[>\!\!10\sigma]$ \\
 & $[0.7,1.0]$ & $0.052_{-0.011}^{+0.013}[7.9\sigma]$ & $0.945_{-0.063}^{+0.061}$ & $4.34_{-0.25}^{+0.26}$ & $0.386_{-0.046}^{+0.050}$ & $0.394_{-0.028}^{+0.027}[>\!\!10\sigma]$ \\
\hline
 & $[0.0,0.4]$ & $0.107_{-0.032}^{+0.056}[7.4\sigma]$ & $1.029_{-0.096}^{+0.098}$ & $4.85_{-0.36}^{+0.36}$ & $0.490_{-0.069}^{+0.073}$ & $0.431_{-0.037}^{+0.041}[>\!\!10\sigma]$ \\
$[600,800]$ & $[0.4,0.7]$ & $0.091_{-0.030}^{+0.044}[5.5\sigma]$ & $0.58_{-0.11}^{+0.11}$ & $3.19_{-0.36}^{+0.25}$ & $0.210_{-0.044}^{+0.048}$ & $0.260_{-0.048}^{+0.050}[8.7\sigma]$ \\
 & $[0.7,1.0]$ & $0.033_{-0.012}^{+0.015}[3.6\sigma]$ & $0.475_{-0.089}^{+0.087}$ & $2.21_{-0.40}^{+0.39}$ & $0.121_{-0.034}^{+0.042}$ & $0.163_{-0.044}^{+0.050}[6.6\sigma]$ \\
\hline
 & $[0.0,0.4]$ & \bm{$0.424_{-0.091}^{+0.078}[9.8\sigma]$} & \bm{$2.03_{-0.15}^{+0.15}[6.7\sigma]$} & \bm{$8.55_{-0.65}^{+0.65}[3.6\sigma]$} & \bm{$0.99_{-0.17}^{+0.20}$} & \bm{$0.561_{-0.066}^{+0.042}[7.3\sigma]$} \\
$[800,13000]$ & $[0.4,0.7]$ & $0.158_{-0.070}^{+0.046}[5.8\sigma]$ & $1.15_{-0.13}^{+0.13}$ & $5.27_{-0.52}^{+0.51}$ & $0.617_{-0.109}^{+0.100}$ & $0.567_{-0.056}^{+0.040}[>\!\!10\sigma]$ \\
 & $[0.7,1.0]$ & $0.033_{-0.015}^{+0.019}[3.1\sigma]$ & $0.169_{-0.088}^{+0.087}$ & $2.01_{-0.40}^{+0.40}$ & $0.071_{-0.027}^{+0.035}$ & $0.113_{-0.037}^{+0.043}[5.0\sigma]$ \\
\hline
\end{tabular}
\label{tab_hel}
\end{table*}

\begin{table*}[htbp]
\renewcommand{\arraystretch}{1.3}
\setlength{\tabcolsep}{5pt}
\centering
\caption{QIS observables in beam basis with their uncertainties, determined by $-2\log(L) = 1$. If this value is not reached for the physically allowed range of the observable, we quote the distance from the boundary, \ie, zero. 
The significance (in units of $\sigma$) of the deviation from the null hypothesis, i.e. \qd$=0$, \DE$=1$, \steer$=2\pi$, \biv$=1$, \Mtwo$=0$, is shown in square brackets when it is greater than $3\sigma$.
}
\begin{tabular}{l l l l l l l}
\hline
\multicolumn{1}{c}{$\ttm$ [GeV]} & \multicolumn{1}{c}{$\absct$} & \multicolumn{1}{c}{\qd} & \multicolumn{1}{c}{\DE} & \multicolumn{1}{c}{\steer} & \multicolumn{1}{c}{\biv} & \multicolumn{1}{c}{\Mtwo} \\
\hline
& $[0.0,0.4]$ & $0.187_{-0.078}^{+0.132}[5.3\sigma]$ & $0.96_{-0.23}^{+0.23}$ & $4.51_{-0.77}^{+0.79}$ & $0.41_{-0.13}^{+0.15}$ & $0.383_{-0.078}^{+0.063}[5.6\sigma]$ \\
$[300,400]$ & $[0.4,0.7]$ & $0.177_{-0.084}^{+0.143}[3.4\sigma]$ & $1.26_{-0.41}^{+0.41}$ & $5.3_{-1.6}^{+1.7}$ & $0.44_{-0.21}^{+0.29}$ & $0.48_{-0.17}^{+0.10}[3.5\sigma]$ \\
 & $[0.7,1.0]$ & $0.376_{-0.083}^{+0.127}[8.2\sigma]$ & $1.88_{-0.28}^{+0.28}[3.2\sigma]$ & $7.9_{-1.1}^{+1.1}$ & $0.94_{-0.21}^{+0.23}$ & $0.543_{-0.085}^{+0.039}[4.6\sigma]$ \\
\hline
 & $[0.0,0.4]$ & $0.209_{-0.026}^{+0.035}[>\!\!10\sigma]$ & $0.360_{-0.034}^{+0.036}$ & $3.93_{-0.22}^{+0.22}$ & $0.211_{-0.026}^{+0.029}$ & $0.329_{-0.027}^{+0.027}[>\!\!10\sigma]$ \\
$[400,600]$ & $[0.4,0.7]$ & $0.127_{-0.027}^{+0.031}[9.9\sigma]$ & $0.452_{-0.066}^{+0.067}$ & $3.30_{-0.27}^{+0.26}$ & $0.186_{-0.030}^{+0.032}$ & $0.251_{-0.032}^{+0.033}[>\!\!10\sigma]$ \\
 & $[0.7,1.0]$ & $0.1101_{-0.0137}^{+0.0092}[>\!\!10\sigma]$ & $1.004_{-0.059}^{+0.059}$ & $4.21_{-0.25}^{+0.25}$ & $0.237_{-0.033}^{+0.037}$ & $0.365_{-0.031}^{+0.030}[>\!\!10\sigma]$ \\
\hline
 & $[0.0,0.4]$ & $0.017_{-0.012}^{+0.019}$ & $0.545_{-0.053}^{+0.052}$ & $3.58_{-0.34}^{+0.35}$ & $0.301_{-0.055}^{+0.060}$ & $0.264_{-0.024}^{+0.019}[9.8\sigma]$ \\
$[600,800]$ & $[0.4,0.7]$ & $0.0132_{-0.0095}^{+0.0160}$ & $0.351_{-0.055}^{+0.056}$ & $2.45_{-0.37}^{+0.37}$ & $0.131_{-0.037}^{+0.042}$ & $0.162_{-0.038}^{+0.038}[6.1\sigma]$ \\
 & $[0.7,1.0]$ & $0.034_{-0.010}^{+0.012}[4.9\sigma]$ & $0.503_{-0.078}^{+0.079}$ & $2.12_{-0.33}^{+0.32}$ & $0.064_{-0.020}^{+0.022}$ & $0.114_{-0.030}^{+0.035}[5.8\sigma]$ \\
\hline
 & $[0.0,0.4]$ & $0.003_{-0.003}^{+0.034}$ & $0.721_{-0.082}^{+0.083}$ & $4.56_{-0.53}^{+0.53}$ & $0.52_{-0.11}^{+0.13}$ & $0.260_{-0.043}^{+0.019}[3.4\sigma]$ \\
$[800,13000]$ & $[0.4,0.7]$ & $0.0004_{-0.0004}^{+0.0093}$ & $0.609_{-0.069}^{+0.067}$ & $3.82_{-0.42}^{+0.43}$ & $0.370_{-0.078}^{+0.087}$ & $0.2686_{-0.0172}^{+0.0074}[8.4\sigma]$ \\
 & $[0.7,1.0]$ & $0.0013_{-0.0013}^{+0.0035}$ & $0.181_{-0.078}^{+0.078}$ & $0.86_{-0.33}^{+0.34}$ & $0.0153_{-0.0096}^{+0.0154}$ & $0.023_{-0.014}^{+0.021}$ \\
 \hline
\end{tabular}
\label{tab_beam}
\end{table*}

\begin{figure*}[t]
    \centering
    \includegraphics[width=0.48\textwidth]{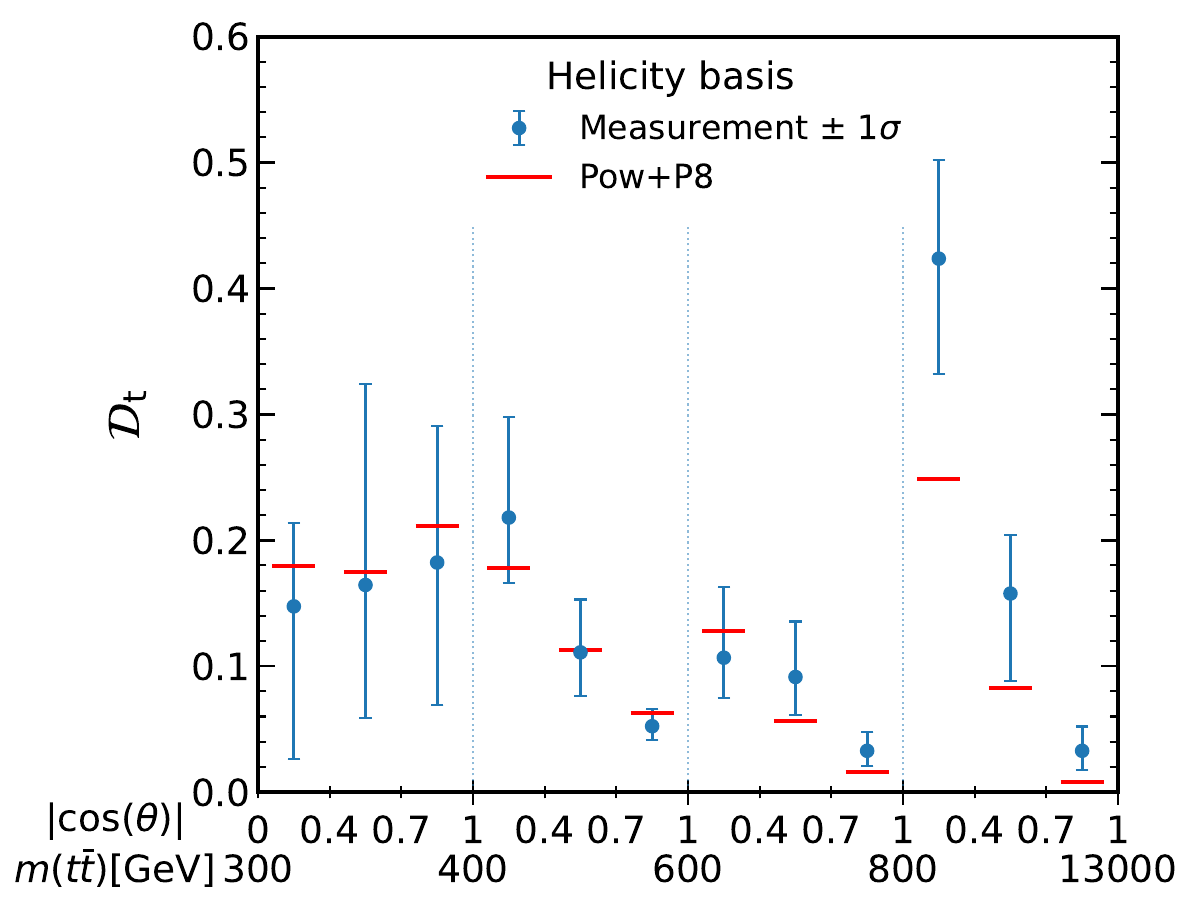}
    \includegraphics[width=0.48\textwidth]{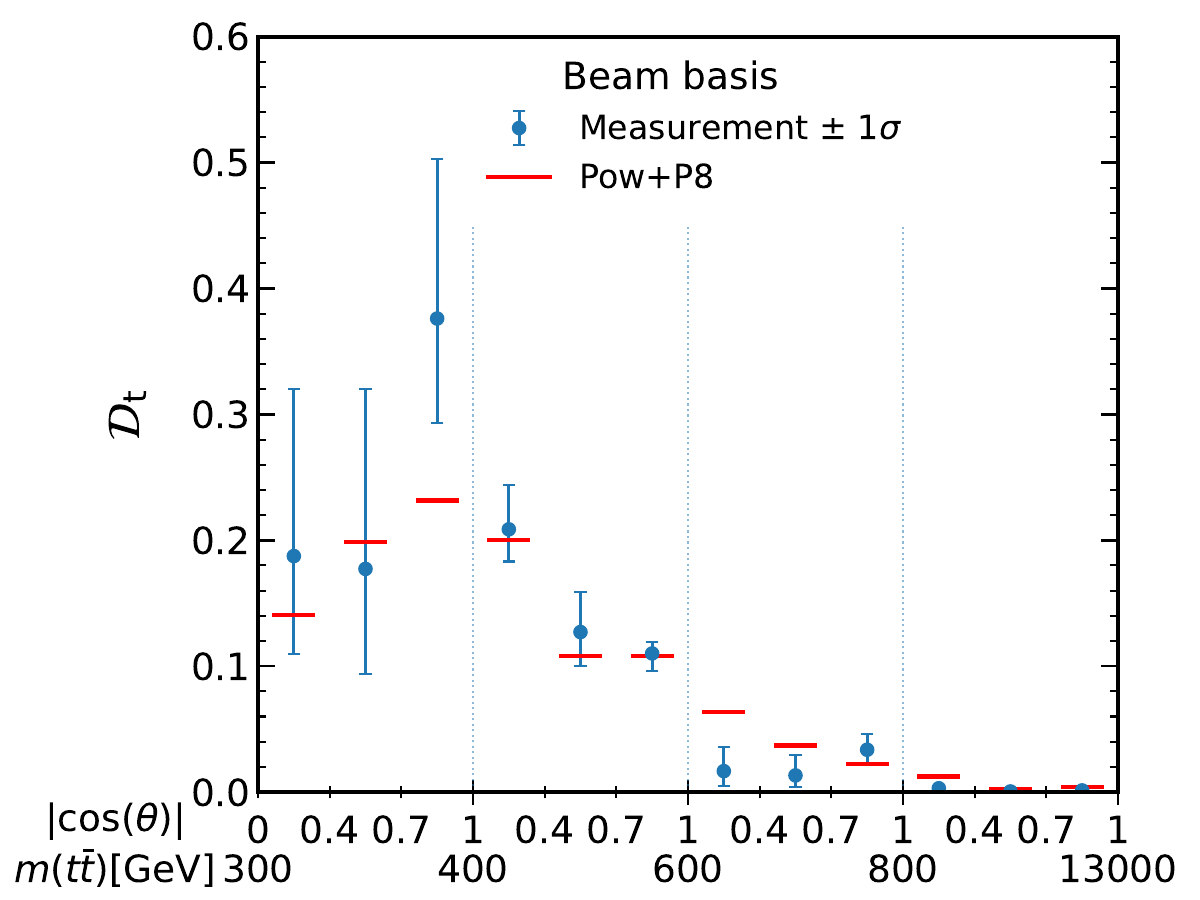}\\
    \includegraphics[width=0.48\textwidth]{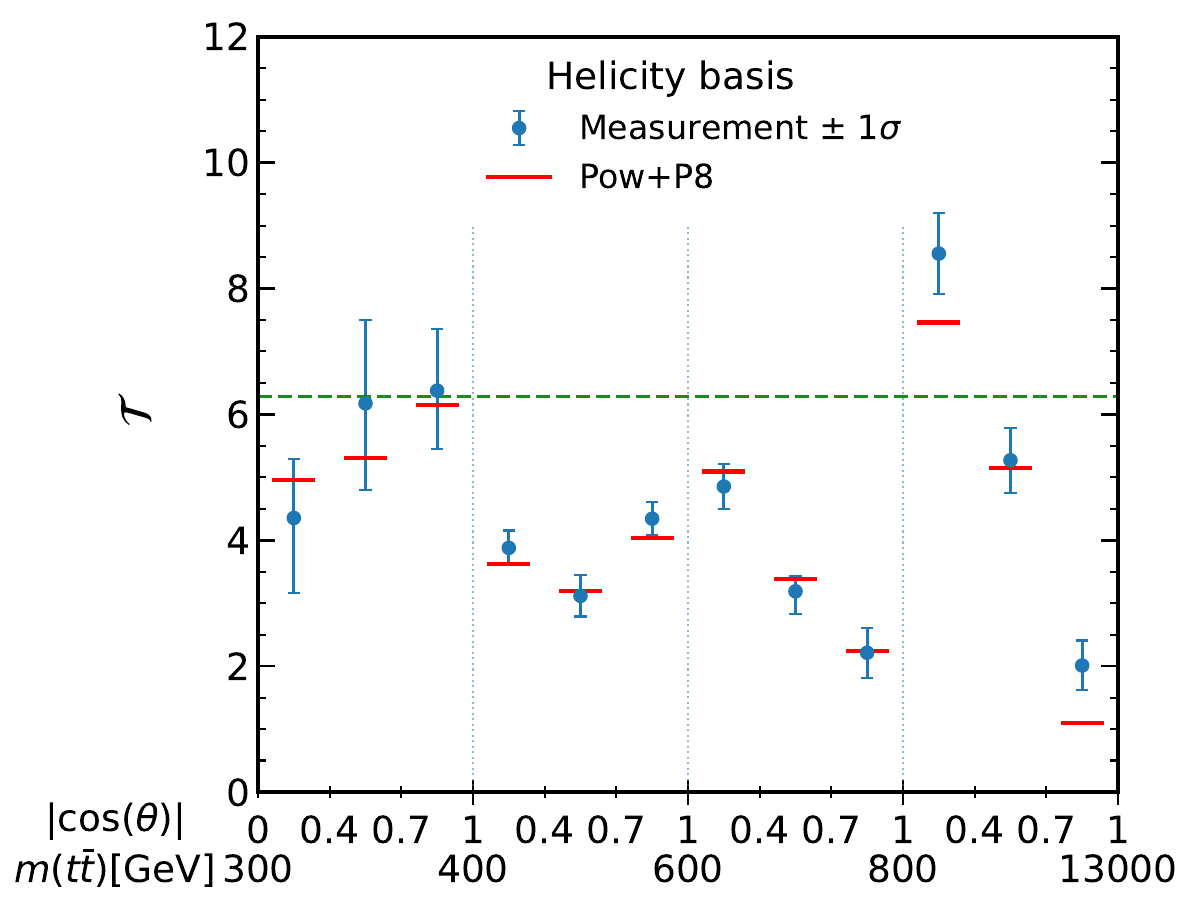}
    \includegraphics[width=0.48\textwidth]{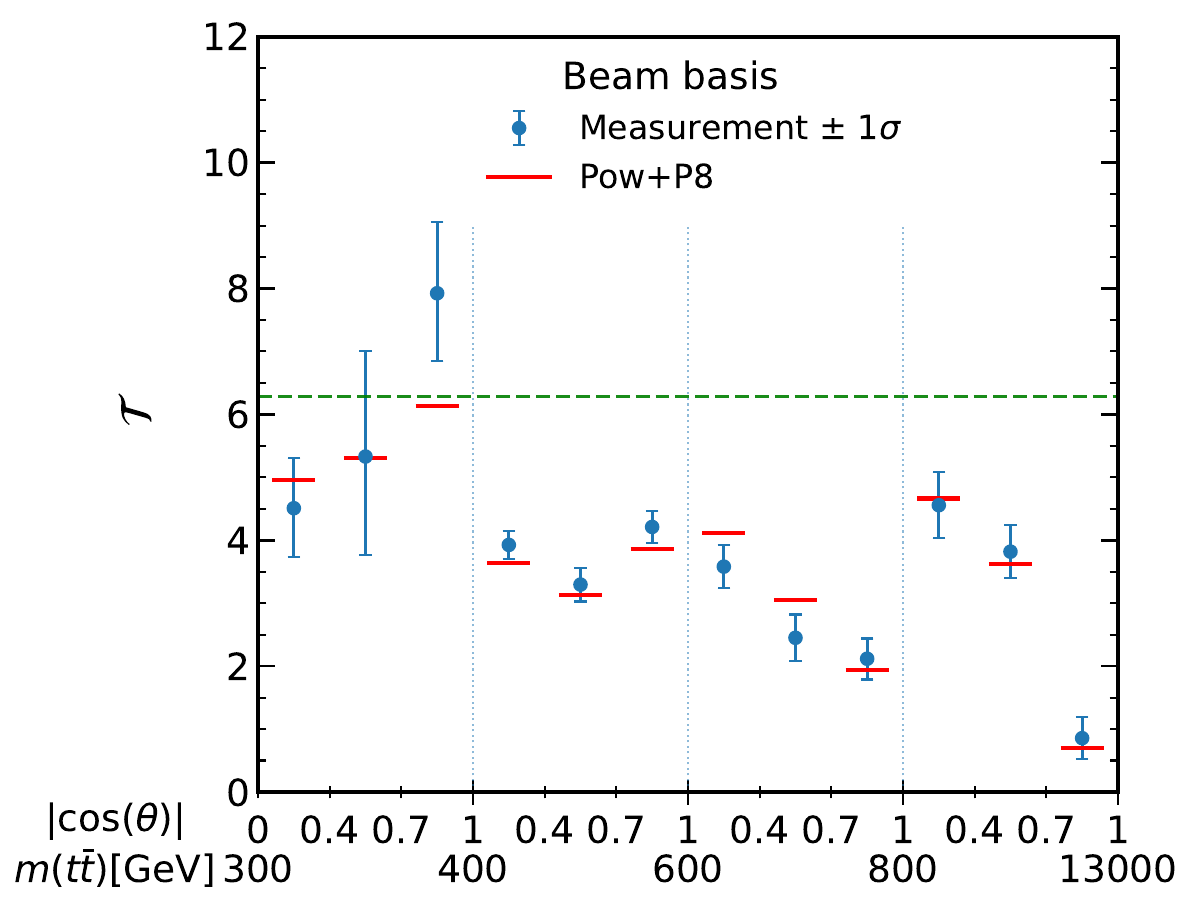}\\
    \includegraphics[width=0.48\textwidth]{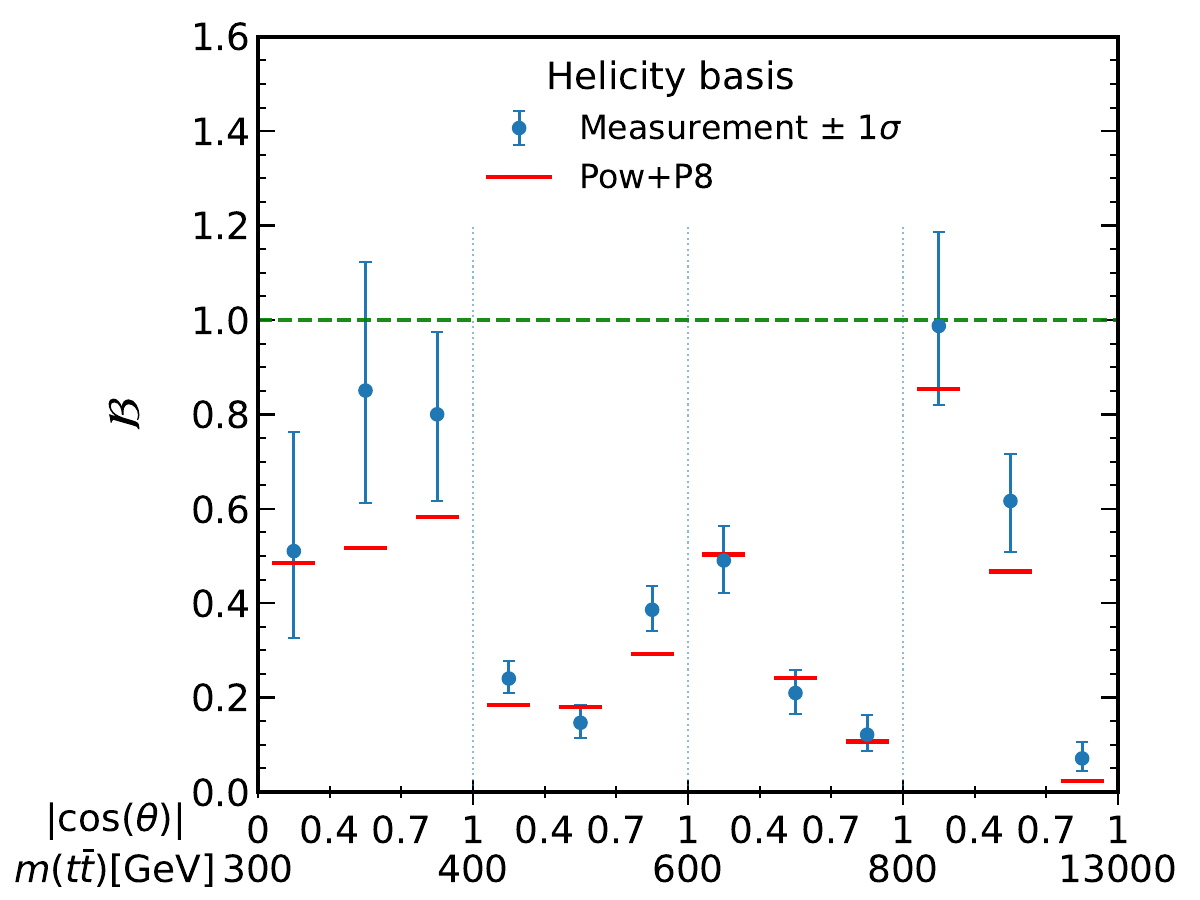}
    \includegraphics[width=0.48\textwidth]{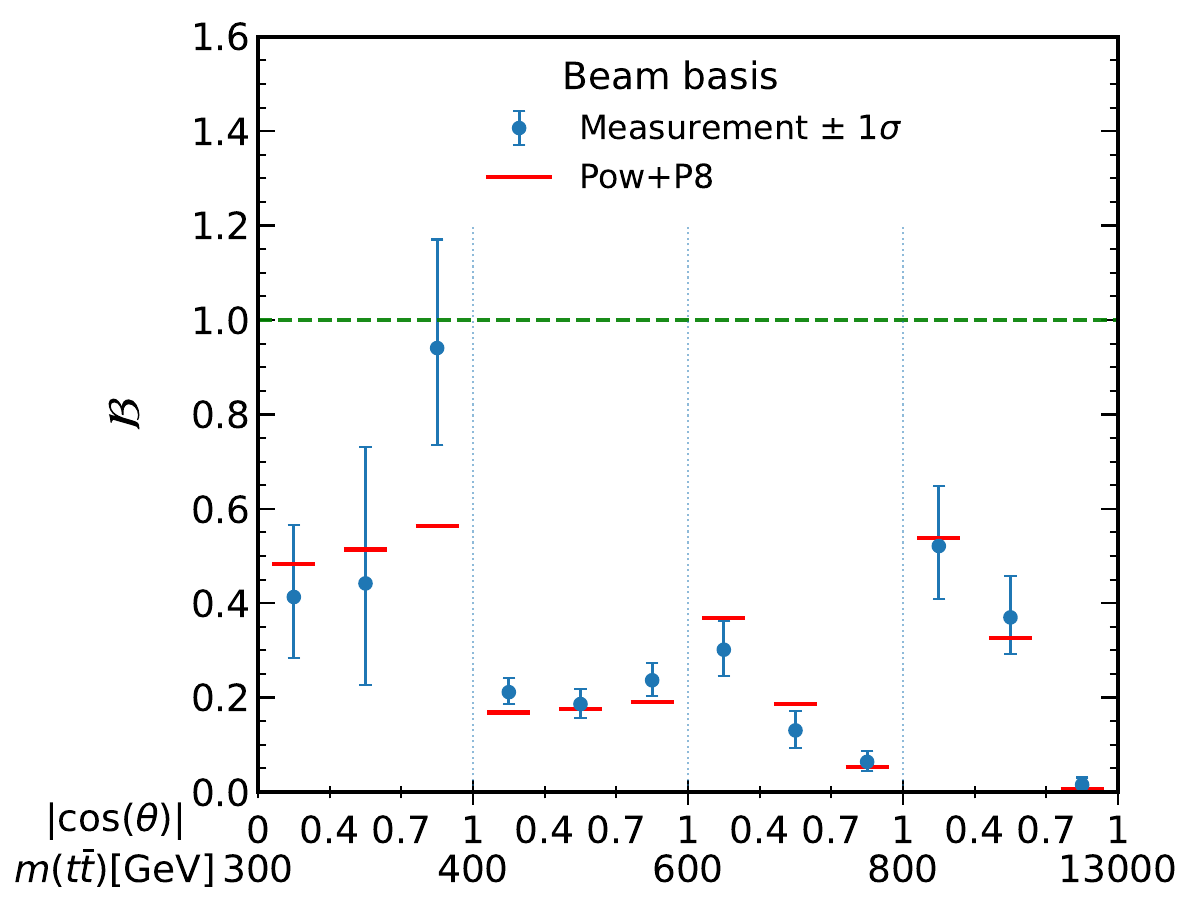} 
    \caption{Results for quantum discord \qd (upper row), steering marker \steer (middle row), and Bell correlation marker \biv (bottom row) in bins of \ttmcth in the helicity (left) and beam (right) bases. The measurements (points) are shown with the total uncertainty and compared to the predictions of \PWGPYT. Horizontal dashed lines indicate the threshold values: 2$\pi$ for steerability and one for Bell correlation. The threshold value for discord is zero.
    }
    \label{resultsdata_dsb}
\end{figure*}

\begin{figure*}
    \centering
    \includegraphics[width=0.48\textwidth]{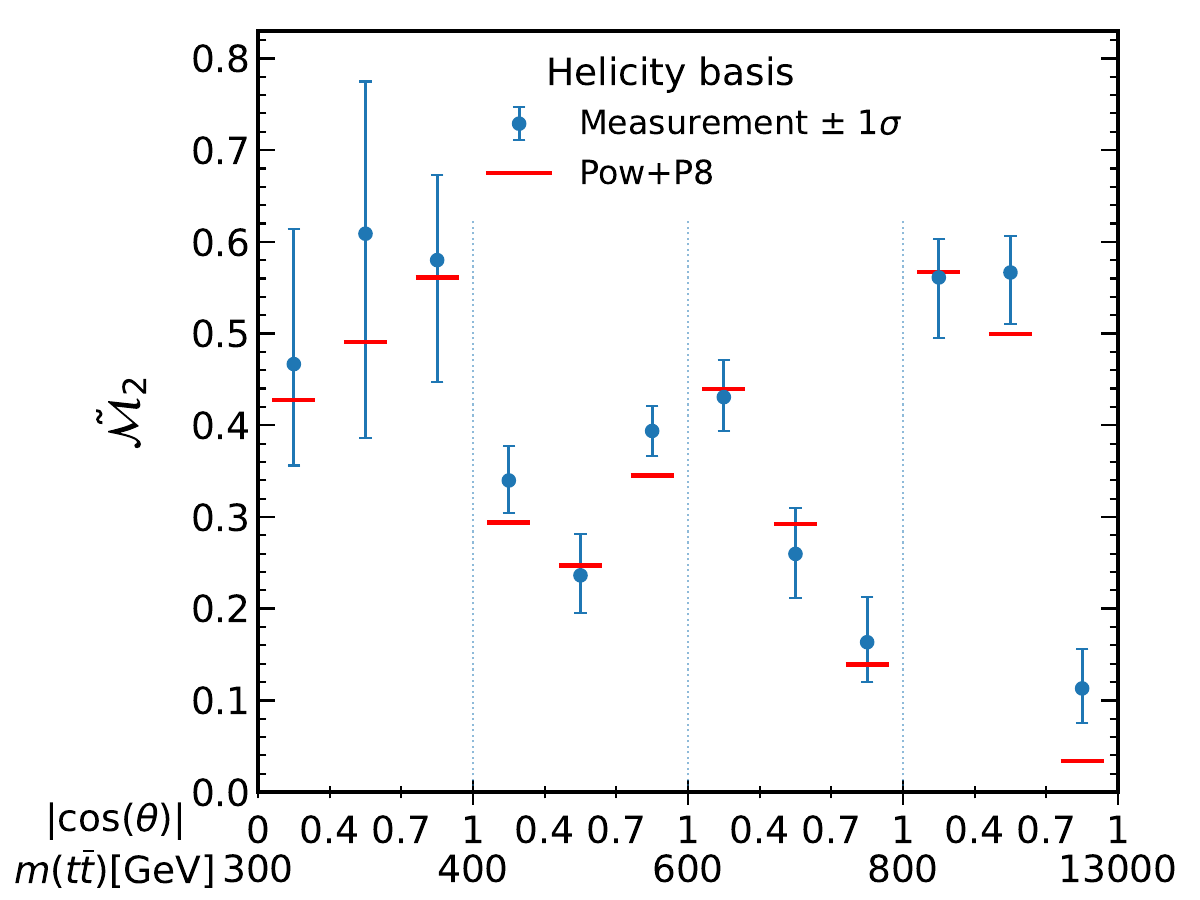}
    \includegraphics[width=0.48\textwidth]{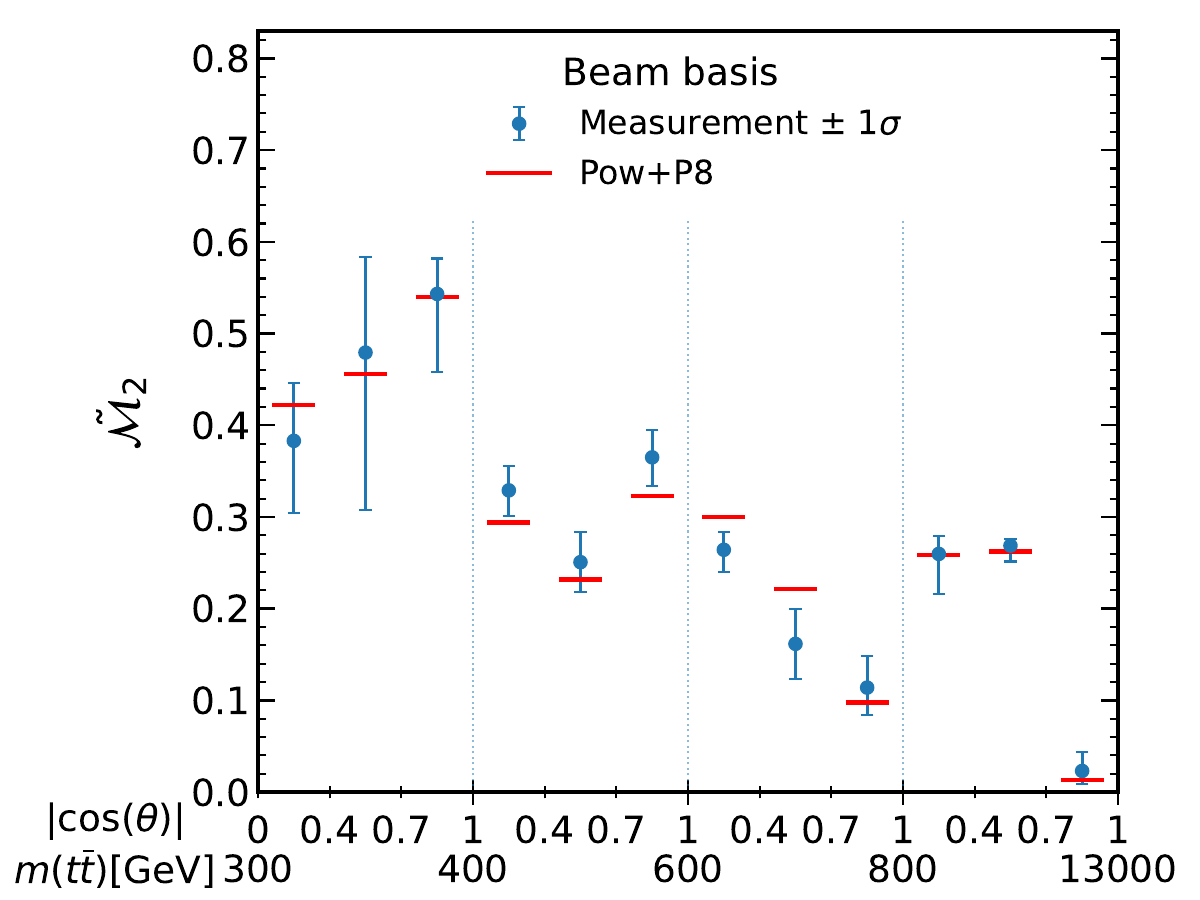}    
    \caption{Quantum magic \Mtwo in the helicity (left) and beam (right) bases in bins of \ttmcth. The measurements (points) are shown with the total uncertainty and compared to the predictions of \PWGPYT. The threshold value for \Mtwo is zero. 
    }
    \label{resultsdata_magic}
\end{figure*}

For the observables shown in Figs.~\ref{resultsdata_dsb} and \ref{resultsdata_magic}, the measured values are compared to the standard model (SM) predictions obtained from the Monte Carlo simulated samples described in Refs.~\cite{CMS:2024zkc,CMS:2025brx}. Since the predictions from the different Monte Carlo simulations employed in these analyses agree well with each other, we include only the one obtained with the \POWHEG{~v2}~\cite{Nason:2004rx,Frixione:2007vw,Frixione:2007nw} matrix-element generator, using \PYTHIA{~8.2}~\cite{Sjostrand:2014zea} for the parton-shower simulation. In general, good agreement is found with the SM predictions for all the measured observables. 

The measurement of the quantum discord of the top quark in the helicity and beam bases in bins of \ttmcth is presented in Fig.~\ref{resultsdata_dsb} (upper row). In the helicity basis, the highest value of discord is observed in the high \ttm and low \absct region, where it exceeds zero with a significance greater than 5$\sigma$. Notably, this region also coincides with that where the most significant entanglement was reported in Ref.~\cite{CMS:2024zkc}. In addition, in several bins we find nonzero discord with $>5\sigma$ significance, as shown in Tables~\ref{tab_hel} and ~\ref{tab_beam}, despite the system being in a separable state ($\DE\leq 1$), which demonstrates the presence of quantum correlations in \ttbar production in the absence of entanglement. In the beam basis, the highest value of discord is obtained in the region with low \ttm and high \absct. In the high \ttm region, discord approaches zero, meaning that the beam basis is not optimal for observing the quantum nature of correlations outside the threshold region.

The definition of discord is not necessarily symmetric under exchanging the top quark and antiquark. For completeness, we present the quantum discord of the top antiquark (\qdb) in Appendix~\ref{app:discord}. As discussed in Ref.~\cite{Afik:2022dgh}, a deviation from zero in the difference between the discord of the top quark and antiquark would signal the presence of $CP$-violating physics beyond the SM. We, therefore, also include in Appendix~\ref{app:discord} the difference $\qd - \qdb$, which we find to be consistent with the SM expectation of zero.

The steerability results for helicity and beam bases in bins of \ttmcth are presented in Fig.~\ref{resultsdata_dsb} (middle row). In the helicity basis, we observe the highest and the most significant value of the steering marker in the $\ttm >800$~GeV, $\absct <0.4$ bin. The observed value of $\steer = 8.55^{+0.65}_{-0.65}$ exceeds the threshold value of 2$\pi$ by more than 3$\sigma$, providing the first evidence for steerability in the \ttbar system. This indicates that, in the phase space region where entanglement has been observed~\cite{CMS:2024zkc}, a higher echelon in the hierarchy of quantum correlations is accessible.

The results for Bell correlations for helicity and beam bases in bins of \ttmcth are presented in Fig.~\ref{resultsdata_dsb} (bottom row). The highest value of \biv 
$=0.99^{+0.20}_{-0.17}$ is observed in the helicity basis in the $\ttm >800$~GeV, $\absct <0.4$ bin, which falls just below the threshold value of 1. 

In Fig.~\ref{resultsdata_magic} the extracted values of magic are shown in bins of \ttmcth for the helicity and beam bases. In the helicity basis the highest \Mtwo value is observed for low \ttm, however, the most significant results are found at central and high \ttm. In the beam basis we observe the highest \Mtwo value at low \ttm and high \absct.

\begin{figure}
    \centering
    \includegraphics[width=0.48\textwidth]{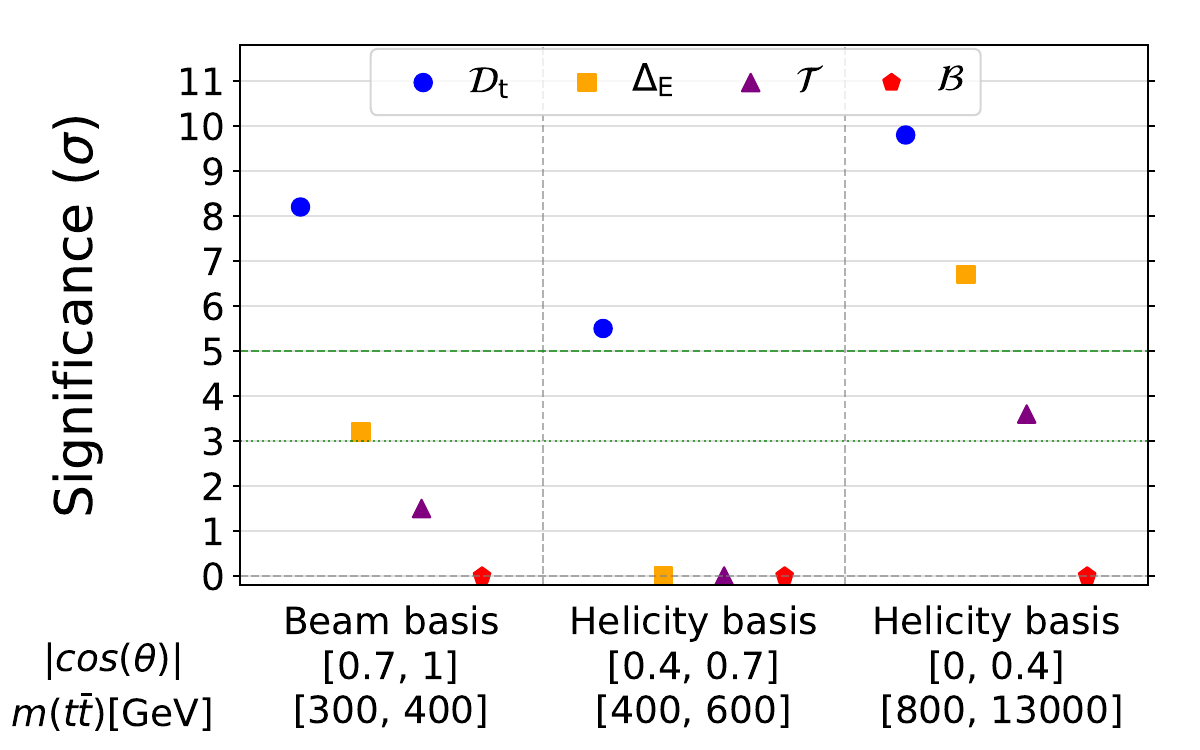}
    \caption{Significance in units of $\sigma$ of the discord, entanglement, steerability and Bell correlation markers exceeding the threshold values. 
    }
    \label{fig:quantum_hierarchy}
\end{figure}

To illustrate the hierarchy of the quantum correlations, we select three bins and plot the significance of the discord, entanglement (from Refs.~\cite{CMS:2024zkc,CMS:2025brx}), steerability, and Bell correlation markers 
exceeding the respective threshold values in Fig.~\ref{fig:quantum_hierarchy}. In the first bin, the hierarchy of the quantum observables is most apparent. The second bin exemplifies the case where a significant value of discord is observed despite \ttbar being in a separable state. In the third bin, \ttbar is observed in an entangled state, the significance of discord exceeds 5$\sigma$, while the significance of steerability is above 3$\sigma$. Bell correlation is not observed in this, or any other bin probed in this study. In Refs.~\cite{Fabbrichesi:2021npl,Severi:2021cnj} it was argued that Bell correlation might be observed in the \ttbar system in the region of phase space with even higher \ttm and lower \absct. Measuring spin correlations in this region with sufficient statistics will require a larger \ttbar sample.

\FloatBarrier

\section{Conclusion}
The production of top quark--antiquark pairs at the LHC offers a unique opportunity to study the fundamental aspects of quantum mechanics in the high-energy regime. In this paper, we present the 
evaluation of a number of quantum information properties, specifically discord,  steerability, Bell correlation, and magic, of the top quark--antiquark system.
The observables are derived from the doubly differential measurements of top quark polarization and spin correlation coefficients published by the CMS collaboration, performed in the helicity and beam bases.
The helicity basis, with its event-dependent orientation, characterizes fictitious quantum states, while the beam basis, with fixed orientation, yields \textit{bona fide} spin quantum states. 
We observe that discord, which constitutes the most basic form of quantum correlations, is greater than zero with a significance of more than 5$\sigma$ in several regions of phase space, some of which correspond to separable quantum states. This observation demonstrates that entanglement is not a necessary condition for quantum behavior of the \ttbar system.
We also find that the proposed steering marker exceeds the threshold value with a significance greater than 3$\sigma$. The threshold value for Bell correlation is not reached. Bell correlations are not expected to arise within the regions of phase space probed in this study.

Notably, the statistical significance of each observable reflects its position in the hierarchy of quantum correlations: tighter correlations are significant only within a smaller region of phase space and, therefore, within a given bin, they exhibit lower statistical significance, as they are more difficult to observe. Magic, which quantifies a potential advantage for quantum computation and is complementary to the quantum-correlation hierarchy, is found to be greater than zero with a significance of more than 5$\sigma$ in several regions of phase space. 

Future measurements, incorporating the larger datasets anticipated from Run~3 of the LHC, are expected to improve the sensitivity to Bell correlations and quantum steering. 
Finally, several studies have shown that quantum correlations constitute a sensitive tool for probing physics beyond the SM~\cite{Aoude:2022imd,Fabbrichesi:2022ovb,Severi:2022qjy,Maltoni:2024tul,Aoude:2025jzc}, and the results presented in this work may contribute to such investigations.

\acknowledgments{AH, OH and RD acknowledge support from the U.S. Department of Energy under the grant DE-SC0008475. YA is supported by the National Science Foundation under Grant No.\ PHY-2310094. JRMdN acknowledges funding from European Union's Horizon 2020 research and innovation programme under the Marie Sk\l{}odowska-Curie Grant Agreement No. 847635, and from Spain's MICIU/AEI through Proyectos de Generaci\'on de Conocimiento (Grant No. PID2022-139288NB-I00) and through Ram\'on y Cajal program (Grant No. RYC2024-050437-I).}

\bibliography{main.bib}

\appendix

\section{More results on discord}
\label{app:discord}
\begin{figure*}
    \centering
    \includegraphics[width=0.48\textwidth]{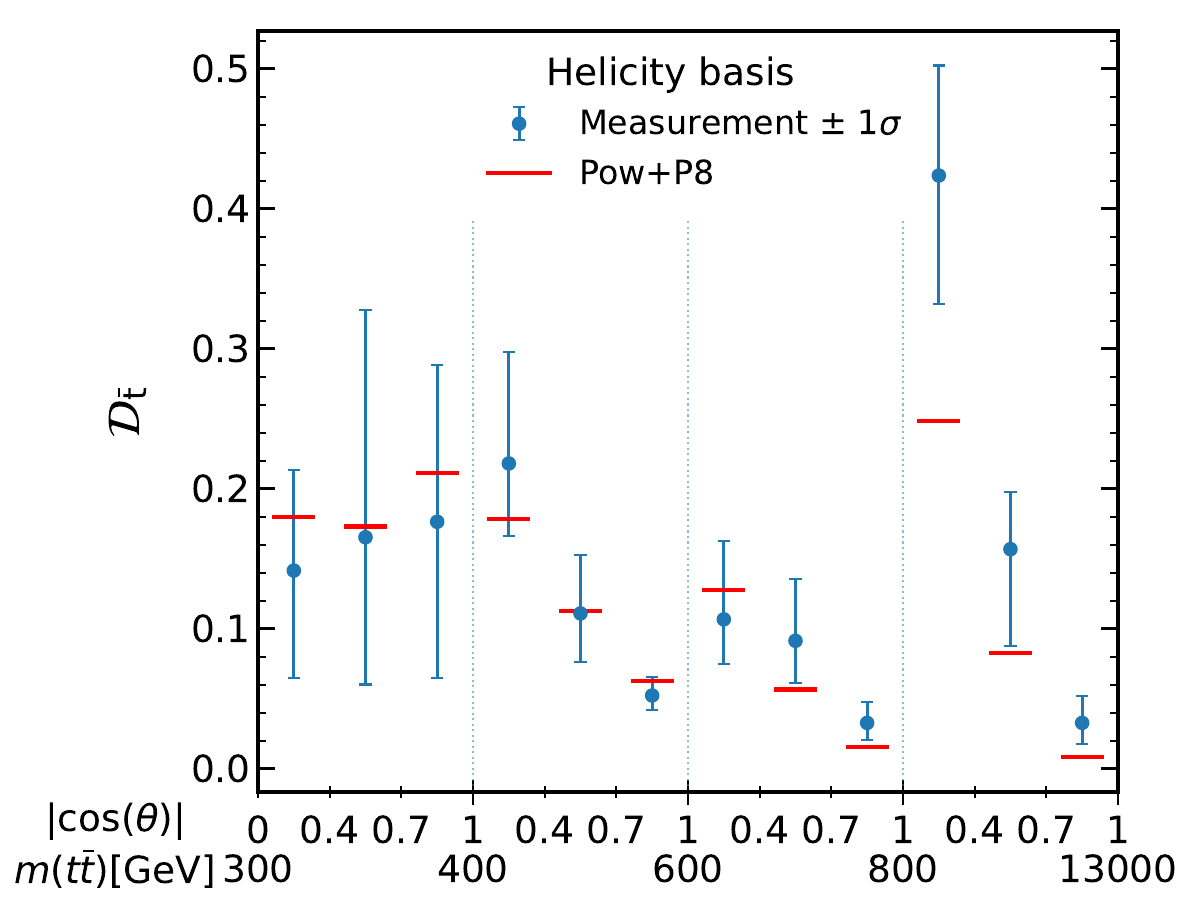}
    \includegraphics[width=0.48\textwidth]{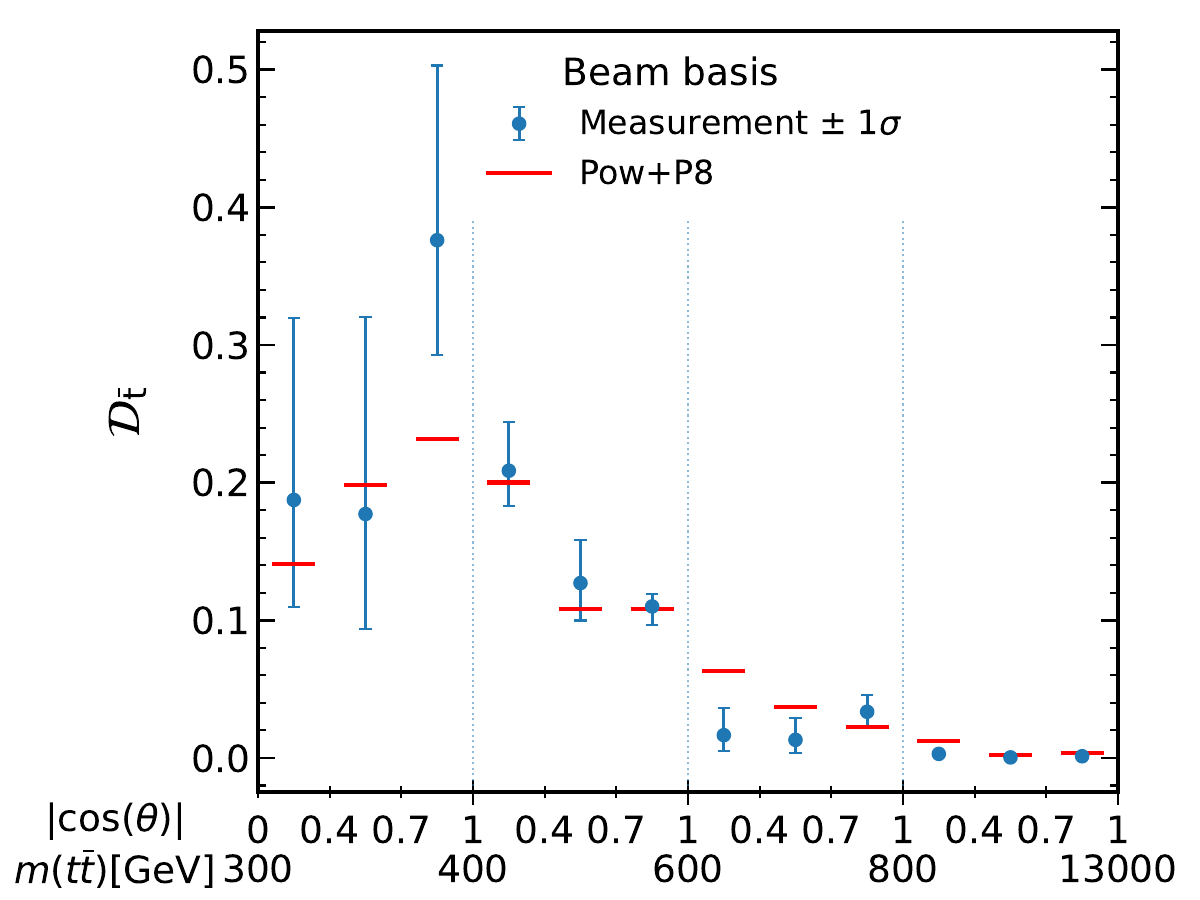}
     \includegraphics[width=0.48\textwidth]{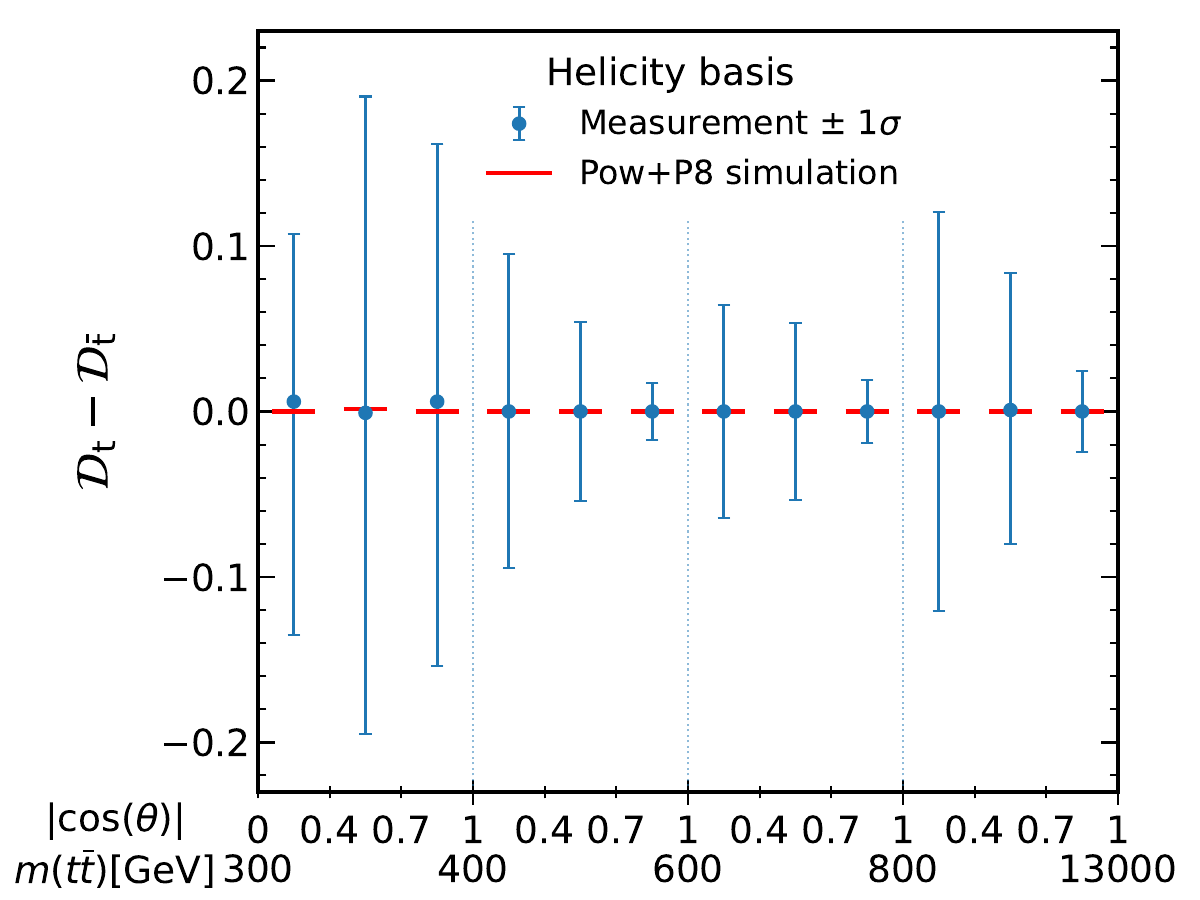}
    \includegraphics[width=0.48\textwidth]{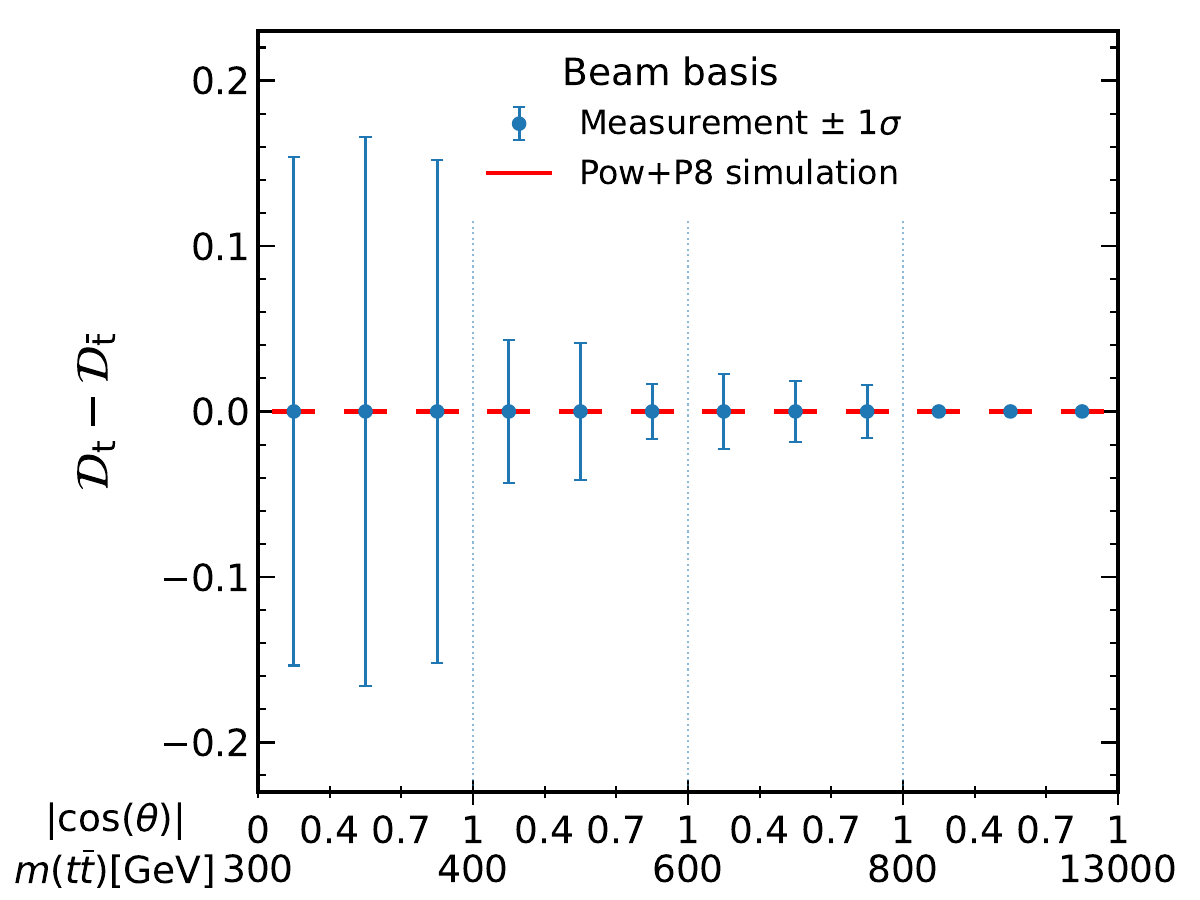}
   \caption{Results of \qdb (upper row) and $\qd-\qdb$ (bottom row) in bins of \ttmcth in the helicity (left) and beam (right) basis. 
    The measurements (points) are shown with the total uncertainty and compared to the predictions of \PWGPYT. The threshold value for discord is zero.}
    \label{resultsdata_discordB}
\end{figure*}

We present the measurements of the top antiquark discord, \qdb, and the difference in the discord of the top quark and antiquark, $\qd-\qdb$, in Fig.~\ref{resultsdata_discordB}.

\end{document}